Санкт-Петербургский Государственный Политехнический Университет

*Физико-Механический Факультет*

*Кафедра Экспериментальной Ядерной Физики*

# Методические указания к решению задач по ядерной физике

Н.И.Троицкая

**Предисловие**

Настоящий сборник методических указаний к решению задач по ядерной физике составлен на основе практических занятий к курсу лекций «Ядерная физика» для студентов 4-х курсов физико-механического факультета по специальности «Техническая физика» и «Медицинская ядерная физика» Санкт-Петербургского Государственного Политехнического Университета. Основная цель сборника – выработать у студентов навыки научно-исследовательского подхода к решению практических задач по ядерной физике.

## 1. Основные характеристики ядер

Задача 10.8. Определить с помощью табличных значений масс нуклидов а) энергию связи нейтрона и α-частицы в ядре $^{21}Ne_{10}$ ;

б) энергию, необходимую для разделения ядра $^{16}O_{8}$ на четыре одинаковые частицы.

Указание. Для решения задач на эту тему напомним обозначения. Ядро **X** (нуклид), содержащее **A** нуклонов и **Z** протонов, обозначают как $^{A}\mathbf{X}_{Z}$ . Для энергии связи ядра $^{A}\mathbf{X}_{Z}$ принято обозначение $\mathbf{E}_{св}(^{A}\mathbf{X}_{Z})$, а для избытка массы – $\Delta(^{A}\mathbf{X}_{Z})$. Избыток массы $\Delta(^{A}\mathbf{X}_{Z})$ нуклида $^{A}\mathbf{X}_{Z}$ определяется как разность $(M(^{A}\mathbf{X}_{Z}) - A)$, где $M(^{A}\mathbf{X}_{Z})$ - наблюдаемая масса ядра, а A – целое число атомных единиц массы (а.е.м.) , сопоставляемых ядру $^{A}\mathbf{X}_{Z}$ . Все величины: $\mathbf{E}_{св}(^{A}\mathbf{X}_{Z})$, $\Delta(^{A}\mathbf{X}_{Z})$, $M(^{A}\mathbf{X}_{Z})$ и A измеряются в а.е.м.- атомных единицах массы. Связь между а.е.м. и МэВ определяется следующим образом

$$1 а.е.м.= 931,494 МэВ/c^2,$$

$$1 а.е.м. \; c^2 = 931,494 МэВ,$$

где $c = 3 \cdot 10^8$ м/сек – скорость света в вакууме. Напомним, что 1 МэВ = $1,602 \cdot 10^{-13}$ Дж.

Энергия связи $\mathbf{E}_{св}(^{A}\mathbf{X}_{Z})$ ядра $^{A}\mathbf{X}_{Z}$ определятся формулой

$$\mathbf{E}_{св}(^{A}\mathbf{X}_{Z}) = Z\,\Delta_H + (A - Z)\,\Delta_n - \Delta(^{A}\mathbf{X}_{Z}),$$

где $\Delta_H$ = 0,007825 а. е. м. и $\Delta_n$ =0,008665 а. е. м. - избыток массы атома водорода и нейтрона, соответственно.

Решение 10.8 а(n).

Величина энергии связи $E_{св}(^{21}Ne_{10})$ ядра $^{21}Ne_{10}$ определяется сильным взаимодействием двадцати одного нуклона в ядре $^{21}Ne_{10}$. Отделяя один нейтрон, мы получим ядро $^{20}Ne_{10}$ с энергией связи $E_{св}(^{20}Ne_{10})$. Таким образом, энергия связи одного нейтрона в ядре $^{21}Ne_{10}$ будет равна разности энергий связи ядра $^{21}Ne_{10}$ и ядра $^{20}Ne_{10}$:

$$E_{св}(n)= E_{св}(^{21}Ne_{10}) - E_{св}(^{20}Ne_{10}), \qquad (1.1a)$$

где $E_{св}(^{21}Ne_{10})$ и $E_{св}(^{20}Ne_{10})$ выражаются через избытки масс протона, нейтрона и ядер $^{21}Ne_{10}$ и $^{20}Ne_{10}$:

$$E_{св}(^{21}Ne_{10})=10\Delta_H + 11\Delta_n - \Delta(^{21}Ne_{10}),$$

$$E_{св}(^{20}Ne_{10})=10\Delta_H +10\Delta_n - \Delta(^{20}Ne_{10}). \qquad (1.2a)$$

Подставим соотношения (1.2а) в (1.1а) и найдём энергию связи одного нуклона в ядре $^{21}Ne_{10}$:

$$E_{св}(n)= \Delta_n - [\Delta(^{21}Ne_{10}) - \Delta(^{20}Ne_{10})]. \qquad (1.3a)$$

Численные значения для избытка масс нейтрона $\Delta_n$, ядра $\Delta(^{21}Ne_{10})$ и ядра $\Delta(^{20}Ne_{10})$

$$\Delta_n = 0.008665 \text{ а. е. м.}$$

$$\Delta(^{21}Ne_{10}) = - 0.006151 \text{ а. е. м.}$$

$$\Delta(^{20}Ne_{10}) = - 0.007560 \text{ а. е. м.} \qquad (1.4a)$$

мы берём из таблицы 7 (стр.205 [1]).

Используя соотношения (1.4а) получим численное значение энергии связи нейтрона в ядре $^{21}Ne_{10}$ :

$$E_{св}(n)= 0.007256 \text{ а. е. м.} \qquad (1.5a)$$

Энергии связи $E_{св}(n)$, измеренная в МэВ, равна:

$$E_{св}(n) = 0.007256 \cdot 931.494 = 6.759 \text{ МэВ}. \qquad (1.6a)$$

**Ответ**: Энергия связи нейтрона в ядре $^{21}Ne_{10}$ равна: $E_{св}(n)$ = 6.759 МэВ.

Решение 10.8 а(α).

Для вычисления энергии связи α – частицы в ядре $^{21}Ne_{10}$ найдем сначала энергию системы 2n2p – системы невзаимодействующих двух нейтронов и двух протонов. Если из ядра $^{21}Ne_{10}$ удалить два нейтрона и два протона, то получим ядро изотопа кислорода $^{17}O_8$ с энергией связи $E_{св}(^{17}O_8)$. Тогда энергия связи системы 2n2p, невзаимодействующих двух нейтронов и двух протонов, в ядре $^{21}Ne_{10}$ будет равна:

$$E_{св}(2n2p)= E_{св}(^{21}Ne_{10})- E_{св}(^{17}O_8)=[\ 10\Delta_H + 11\Delta_n - \Delta(^{21}Ne_{10})]-$$

$$-\ [\ 8\Delta_H + 9\Delta_n - \Delta(^{17}O_8)]= 2\Delta_H +2\Delta_n -[\ \Delta(^{21}Ne_{10}) - \Delta(^{17}O_8)]. \qquad (1.7а)$$

Напомним, что $\alpha$ – частица это ядро $^4He_2$, т. е. связанное состояние системы взаимодействующих двух нейтронов и двух протонов 2n2p. Следовательно, для определения энергии связи $\alpha$ – частицы в ядре $^{21}Ne_{10}$ нам надо вычесть энергию связи $E_{св}(^4He_2)$ из $E_{св}(2n2p)$ – энергии связи системы невзаимодействующих двух нейтронов и двух протонов в ядре $^{21}Ne_{10}$:

$$E_{св}(\alpha)= E_{св}(2n2p)- E_{св}(^4He_2)= 2\Delta_H + 2\Delta_n - [\Delta\ (^{21}Ne_{10}) - \Delta(^{17}O_8)] - [2\Delta_H +$$

$$+2\Delta_n - \Delta\ (^4He_2)\ ]= \Delta(^4He_2) +\Delta(^{17}O_8) - \Delta(^{21}Ne_{10}). \qquad (1.8а)$$

Используя численные значения для избытка масс гелия $\Delta\ (^4He_2)$, изотопа кислорода $\Delta(^{17}O_8)$ и ядра $\Delta(^{20}Ne_{10})$ (см. Таблица 7, стр. 205 [1]), равные:

$$\Delta\ (^4He_2) = 0.002604 \text{ а. е. м.}$$

$$\Delta\ (^{17}O_8) = -\ 0.000867 \text{ а. е. м.}$$

$$\Delta\ (^2Ne_{10}) = -\ 0.006\ 151 \text{ а. е. м.}, \qquad (1.9а)$$

получим энергию связи $\alpha$ – частицы ядре $^{21}Ne_{10}$:

$$E_{св}\ (\alpha) = 0.007888 \text{ а. е. м.} \qquad (1.10а)$$

Энергия связи $\alpha$ – частицы $E_{св}\ (\alpha)$, измеренная в МэВ, равна:

$$E_{св}\ (\alpha) = 0.007888*931{,}494 = 7{,}348 \text{ МэВ}. \qquad (1.11а)$$

**Ответ**: Энергия связи $\alpha$ - частицы в ядре $^{21}Ne_{10}$ равна $E_{св}(\alpha)$ = 7,348 МэВ.

Решение 10.8 б.

Определим сначала тип частицы $^AX_Z$. По условию задачи кислород $^{16}O_8$ содержит четыре частицы $^AX_Z$, т. е.

$$\begin{cases} 16 = 4A, \\ 8 = 4Z. \end{cases} \qquad (1.1б)$$

Решение этой системы алгебраических уравнений дает A = 4 и Z = 2.

Отсюда следует, что частица $^4X_2$ является ядром гелия $^4He_2$ или α – частицей. Для определения энергии связи четырех α – частиц в ядре кислорода $^{16}O_8$ надо из энергии связи кислорода исключить энергии связи четырех ядер гелия:

$$E_{св}(4\alpha) = E_{св}(^{16}O_8) - 4E_{св}(^4He_2) =$$

$$= [8\Delta_H + 8\Delta_n - \Delta(^{16}O_8)] - 4[2\Delta_H + 2\Delta_n - \Delta(^4He_2)] = 4\Delta(^4He_2) - \Delta(^{16}O_8). \qquad (1.2б)$$

Используя численные значения для избытка масс ядра гелия $\Delta\ (^4He_2)$ и изотопа кислорода $\Delta\ (^{17}O_8)$ (см. таблицу 7, стр. 205 [1]) равные:

$$\Delta(^4He_2) = 0.002604 \text{ а.е.м.}$$

$$\Delta(^{16}O_8) = -0.005085 \text{ а.е.м.}, \qquad (1.3б)$$

получим энергию связи четырёх α – частиц в ядре кислорода $^{16}O_8$:

$$E_{св}(4\alpha) = 0.015501 \text{ а.е.м.} \qquad (1.4б)$$

Энергия связи четырёх α – частиц в ядре кислорода $^{16}O_8$, измеренная в МэВ, равна:

$$E_{св}(4\alpha) = 0.015501 \cdot 931.494 = 14.439 \text{ МэВ}. \qquad (1.5б)$$

**Ответ**: Энергия связи четырёх α-частиц в ядре кислорода $^{16}O_8$ равна $E_{св}(4\alpha)$ = 14.439 МэВ. Следовательно, для разделения ядра кислорода $^{16}O_8$ на четыре α-частицы,

$$^{16}O_8 \rightarrow 4\ \alpha, \qquad (1.6б)$$

надо приложить энергию E большую или равную энергии связи $E_{св}(4\alpha)$, т.е. $E \geq E_{св}(4\alpha) = 14.439$ МэВ.

## 2. Радиоактивность

Задача 11.13. При радиоактивном распаде ядер нуклида $A_1$ образуется радионуклид $A_2$. Их постоянные распада равны $\lambda_1$ и $\lambda_2$, соответственно. Полагая, что в начальный момент времени препарат содержит только ядра нуклида $A_1$ в количестве $N_{10}$, определить:

а) количество ядер $A_2$ через промежуток времени t;

б) промежуток времени, через который количество ядер $A_2$ достигнет максимума;

в) в каком случае может возникнуть состояние переходного равновесия, при котором отношение количества обоих нуклидов будет оставаться постоянным. Чему равно это отношение?

Указания. Все распады нуклидов происходят по основному закону:

$$N(t) = N_0(t)e^{-\lambda t}\ ,\ \lambda = \frac{1}{\tau} = \frac{\ln 2}{T_{1/2}},$$

где постоянная распада λ определяется средним временем жизни радионуклида τ или периодом полураспада нуклида $T_{1/2}$.

Скорость распада нуклида (т.е. число распадов в единицу времени) определяется уравнением:

$$\frac{dN(t)}{dt} = -\lambda N(t),$$

где $\lambda N(t)$ называют активностью радиоактивного источника и обозначают

$$A(t) = \lambda N(t).$$

В системе СИ активность измеряется в беккерелях (Бк) [2]:
1 Бк = 1/сек, т.е. один беккерель соответствует одному распаду в секунду.
Напомним, что до 1980 года активность измерялась в кюри: 1(Ки) = $3.7\cdot 10^{10}$Бк [2].

Решение 11.13 а.

В соответствии с основным законом радиоактивного распада скорость распада радионуклида $A_1$ в момент времени t (или активность радионуклида в момент времени t) равна:

$$\frac{dN_1(t)}{dt} = -\lambda_1 N_1(t), \qquad (2.1а)$$

где $N_1(t)$ – число радионуклидов $A_1$ в момент времени t.

Уравнение (2.1а) удобно переписать в виде:

$$\frac{dN_1(t)}{dt} + \lambda_1 N_1(t) = 0. \qquad (2.2а)$$

Метод решения уравнения (2.2а), который мы изложим ниже, будет очень удобен для решения многоступенчатых распадов радионуклидов.

Умножим обе части равенства (2.2а) на $e^{\lambda_1 t}$:

$$\frac{dN_1(t)}{dt}e^{\lambda_1 t} + \lambda_1 N_1(t)e^{\lambda_1 t} = 0. \qquad (2.3а)$$

Нетрудно видеть, что левая часть уравнения (2.3а) является полной производной по времени:

$$\frac{d}{dt}\left(N_1(t)e^{\lambda_1 t}\right) = 0. \qquad (2.4а)$$

Из уравнения (2.4а) получаем:

$$N_1(t)e^{\lambda_1 t} = C_1, \qquad (2.5а)$$

где $C_1$ – константа, которая определяется из начальных условий.

Решая уравнение (2.5а) относительно $N_1(t)$, получаем:

$$N_1(t) = C_1 e^{-\lambda_1 t}. \qquad (2.6а)$$

Константу $C_1$ находим из начального условия:

$$N_1(0) = N_{10}. \qquad (2.7а)$$

Полагая в (2.6а) $t = 0$ и используя начальное условие (2.7а), выразим константу $C_1$ через $N_{10}$:

$$N_1(0) = C_1 = N_{10}. \qquad (2.8а)$$

Таким образом, находим, что распад радионуклида $A_1$ происходит по закону:

$$N_1(t) = N_{10}e^{-\lambda_1 t}. \qquad (2.9а)$$

Перейдём к определению скорости распада (или активности) радионуклида $A_2$.

В соответствии с основным законом радиоактивного распада, скорость изменения числа радионуклидов $A_2$ в момент времени t равна:

$$\frac{dN_2(t)}{dt} = \left(\frac{dN_2(t)}{dt}\right)_{образ.} + \left(\frac{dN_2(t)}{dt}\right)_{распада}, \qquad (2.10а)$$

где первое слагаемое определяет скорость образования радионуклида $A_2$ в момент времени t, а второе – скорость распада радионуклида $A_2$ в момент времени t. Скорость образования радионуклида $A_2$ определяется активностью радионуклида $A_1$ и равна:

$$\left(\frac{dN_2(t)}{dt}\right)_{образ.} = \lambda_1 N_1(t). \qquad (2.11а)$$

Скорость распада радионуклида $A_2$ определяется основным законом радиоактивного распада и равна:

$$\left(\frac{dN_2(t)}{dt}\right)_{распада} = -\lambda_2 N_2(t). \qquad (2.12а)$$

Подставляя (2.11а) и (2.12а) в (2.10а) получаем уравнение для скорости распада радионуклида $A_2$:

$$\frac{dN_2(t)}{dt} = \lambda_1 N_1(t) - \lambda_2 N_2(t). \qquad (2.13а)$$

Поскольку в начальный момент времени радионуклид $A_2$ отсутствовал, то уравнение (2.13а) решаем при начальном условии:

$$N_2(t)|_{t=0} = 0. \qquad (2.14а)$$

Для решения уравнения (2.13а) используем метод, изложенный выше. Помня, что $N_1(t) = N_{10}e^{-\lambda_1 t}$, перепишем уравнение (2.13а) в виде:

$$\frac{dN_2(t)}{dt} + \lambda_2 N_2(t) = \lambda_1 N_{10} e^{-\lambda_1 t}. \qquad (2.15а)$$

Умножим обе части на $e^{\lambda_2 t}$:

$$\frac{d}{dt}\left(N_2(t)e^{\lambda_2 t}\right) = \lambda_1 N_{10} e^{(\lambda_2-\lambda_1)\, t} \qquad (2.16а)$$

и проинтегрируем это равенство по времени:

$$N_2(t)e^{\lambda_2 t} = \frac{\lambda_1}{\lambda_2-\lambda_1} N_{10} e^{(\lambda_2-\lambda_1)\, t} + C_2, \qquad (2.17а)$$

где $C_2$ – постоянная интегрирования, определяемая из начальных условий.

Умножая обе части на $e^{-\lambda_2 t}$

$$N_2(t) = \frac{\lambda_1}{\lambda_2-\lambda_1} N_{10} e^{-\lambda_1 t} + C_2 e^{-\lambda_2 t}, \qquad (2.18а)$$

получаем число радионуклидов $A_2$ к моменту времени t.

С учётом начального условия (2.14а) определим константу $C_2$:

$$C_2 = -\frac{\lambda_1}{\lambda_2 - \lambda_1} N_{10}\,. \qquad (2.19а)$$

Таким образом, число радионуклидов в момент времени t определяется соотношением

$$N_2(t) = \frac{\lambda_1}{\lambda_2 - \lambda_1} N_{10}\left(e^{-\lambda_1 t} - e^{-\lambda_2 t}\right). \qquad (2.20а)$$

**Ответ**: Количество ядер $A_2$ через промежуток времени t окажется равным

$$N_2(t) = \frac{\lambda_1}{\lambda_2 - \lambda_1} N_{10}\left(e^{-\lambda_1 t} - e^{-\lambda_2 t}\right).$$

Решение 11.13 б.

Определение промежутка времени, через который количество ядер $A_2$ достигнет максимума, сводится к математической задаче исследования экстремума функции $N_2(t)$, заданной уравнением (2.20а).

Положение экстремума определяется корнями уравнения:

$$\frac{dN_2(t)}{dt} = 0\,. \qquad (2.1б)$$

Дифференцируя $N_2(t)$ в (2.20а) по t и приравнивая производную нулю, получим:

$$\frac{dN_2(t)}{dt} = \frac{\lambda_1}{\lambda_2 - \lambda_1} N_{10}\left(-\lambda_1 e^{-\lambda_1 t} + \lambda_2 e^{-\lambda_2 t}\right) = 0\,. \qquad (2.2б)$$

Это уравнение имеет только один корень:

$$t_m = \frac{1}{\lambda_1 - \lambda_2} \ln\frac{\lambda_1}{\lambda_2}\,. \qquad (2.3б)$$

Таким образом, функция $N_2(t)$, заданная уравнением (2.20а), имеет только один экстремум в точке $\boldsymbol{t_m}$.

Чтобы определить свойства этого экстремума надо, найти вторую производную в точке $\boldsymbol{t_m}$ .

$$\frac{d^2N_2(t)}{dt^2}\Big|_{t=t_m} = \frac{\lambda_1}{\lambda_2-\lambda_1}N_{10}\left(\lambda_1{}^2e^{-\lambda_1 t_m} - \lambda_2{}^2e^{-\lambda_2 t_m}\right) =$$
$$= \frac{\lambda_1}{\lambda_2-\lambda_1}N_{10}e^{-\lambda_2 t_m}\left(\lambda_1{}^2e^{(\lambda_2-\lambda_1)t_m} - \lambda_2{}^2\right) = -\lambda_1\lambda_2N_{10}e^{-\lambda_2 t_m} < 0 \qquad (2.4б)$$

Отрицательный знак второй производной свидетельствует о том, что в точке $t = t_m$, где $t_m$ определена уравнением (2.3б), функция $N_2(t)$, заданная уравнением (2.20а), имеет максимум.

**Ответ**: Количество ядер нуклида $A_2$ достигает максимума через промежуток времени $t = t_m = \frac{1}{\lambda_1-\lambda_2}\ln\frac{\lambda_1}{\lambda_2}$ с начала распада нуклида $A_1$.

Решение 11.3 в.

Состояние переходного равновесия определяется уравнением

$$\frac{N_2(t)}{N_1(t)} = const. \qquad (2.1в)$$

Числа $N_1(t)$ и $N_2(t)$ нуклидов $A_1$ и $A_2$, соответственно, равны:

$$N_1(t) = N_{10}e^{-\lambda_1 t},$$
$$N_2(t) = \frac{\lambda_1}{\lambda_2-\lambda_1}N_{10}\left(e^{-\lambda_1 t} - e^{-\lambda_2 t}\right), \qquad (2.2в)$$

(см. уравнения (2.9а) и (2.20а)).

Подставим числа нуклидов, заданные уравнением (2.2в), в уравнение (2.1в) и получаем:

$$\frac{N_2(t)}{N_1(t)} = \frac{\lambda_1}{\lambda_2-\lambda_1}\left(1 - e^{-(\lambda_2-\lambda_1)t}\right). \qquad (2.3в)$$

Отсюда видно, что временная зависимость отношения становится практически несущественной, если $\lambda_2 \gg \lambda_1$, т.е. период полураспада нуклида $A_1$ значительно больше периода полураспада нуклида $A_2$, $T_{1/2}(A_1) \gg T_{1/2}(A_2)$. В этом случае в отношении (2.3в) можно пренебречь $e^{-(\lambda_2-\lambda_1)t}$ по сравнению с единицей

$$\frac{N_2(t)}{N_1(t)} \approx \frac{\lambda_1}{\lambda_2} \approx \frac{T_{1/2}(A_2)}{T_{1/2}(A_1)}, \qquad (2.4в)$$

где мы пренебрегли также $\lambda_1$ по сравнению с $\lambda_2$.

**Ответ**: Состояние переходного равновесия в распаде нуклида $A_1 \to A_2$ может возникнуть, если период полураспада нуклида $A_1$ значительно больше периода полураспада нуклида $A_2$, $T_{1/2}(A_1) >> T_{1/2}(A_2)$:

$$\frac{N_2(t)}{N_1(t)} \approx \frac{T_{1/2}(A_2)}{T_{1/2}(A_1)}.$$

## 3. α-распады ядер

Задача 11.25. Покоящееся ядро $^{213}Po_{84}$ испустило α-частицу с кинетической энергией $T_\alpha$ = 8.34 МэВ. При этом дочернее ядро оказалось непосредственно в основном состоянии. Найти полную энергию, высвободившуюся в этом процессе. Какую долю этой энергии составляет кинетическая энергия дочернего ядра? Какая скорость отдачи дочернего ядра?

Указание. Для ответа на вопросы задачи надо сначала определить дочернее ядро в реакции: $^{213}Po_{84} \to {}^{A}X_{Z} + \alpha$.

Поскольку α-частица содержит 2 нейтрона и 2 протона, то числа A и Z должны удовлетворять уравнениям:

$$213 = A + 4$$

$$84 = Z + 2.$$

Решая эту систему линейных алгебраических уравнений, получим:

$$A = 209, \quad Z = 82.$$

Таким образом, дочернее ядро имеет 209 нуклонов и 82 протона, следовательно, это свинец – $^{209}Pb_{82}$. Стало быть, реакция, которую мы изучаем, имеет вид:

$$^{213}Po_{84} \to {}^{209}Pb_{82} + \alpha.$$

Энергию, освобождаемую в ядерной реакции, обозначают буквой ***Q*** и называют энергией реакции. По определению, энергия реакции равна разности энергий покоя всех частиц в начальном состоянии и всех частиц в конечном состоянии исследуемой ядерной реакции. Используя закон сохранения энергии, энергию реакции ***Q*** можно выразить также через кинетические энергии частиц ядерной реакции.

Решение 11.25 а. В соответствии с определением, для реакции $^{213}Po_{84} \rightarrow {}^{209}Pb_{82} + \alpha$ энергия реакции ***Q*** равна:

$$\boldsymbol{Q} = [M(^{213}Po_{84}) - M(^{209}Pb_{82}) - M_\alpha]c^2, \qquad (3.1а)$$

где

$$M(^{213}Po_{84}) = 212.993 \text{ а.е.м.}$$
$$M(^{209}Pb_{82}) = 208.981 \text{ а.е.м.} \qquad (3.2а)$$
$$M_\alpha = M(^4He_2) = 4.003 \text{ а.е.м.}$$

- массы взаимодействующих ядер.

Подставив численные значения (3.2а) в (3.1а), вычислим энергию, освобождаемую в реакции $^{213}Po_{84} \rightarrow {}^{209}Pb_{82} + \alpha$:

$$\boldsymbol{Q} = 0.009 \text{ а.е.м.} \qquad (3.3а)$$

Энергия реакции, измеренная в МэВ, равна:

$$\boldsymbol{Q} = 0.009 \cdot 931.494 = 8.384 \text{ МэВ.} \qquad (3.4а)$$

**Ответ**: Энергия, освобождаемая в реакции $^{213}Po_{84} \rightarrow {}^{209}Pb_{82} + \alpha$, равна ***Q*** = 8.384 МэВ.

Решение 11.25 б. Для вычисления доли кинетической энергии дочернего ядра $^{209}Pb_{82}$ в энергии реакции ***Q*** в реакции $^{213}Po_{84} \rightarrow {}^{209}Pb_{82} + \alpha$ надо воспользоваться законом сохранения энергии и импульса.

$$M(^{213}Po_{84})c^2 = M(^{209}Pb_{82})c^2 + T_{Pb} + M_\alpha c^2 + T_\alpha, \qquad (3.1б)$$

где $E_{Po} = M(^{213}Po_{84})c^2$ – полная энергия начального ядра, а $E_{Pb} = M(^{209}Pb_{82})c^2 + T_{Pb}$ и $E_\alpha = M_\alpha c^2 + T_\alpha$ – полные энергии ядер $^{209}Pb_{82}$ и α, а $T_{Pb}$ и $T_\alpha$ – их кинетические энергии.

Поскольку ядро $^{213}Po_{84}$ покоится, то его импульс равен нулю. В этом случае закон сохранения импульса в ядерной реакции $^{213}Po_{84} \rightarrow {}^{209}Pb_{82} + \alpha$ имеет вид:

$$\vec{O} = \vec{P}_{Pb} + \vec{P}_\alpha, \qquad (3.2б)$$

где $\vec{P}_{Pb}$ и $\vec{P}_\alpha$ - импульсы дочернего ядра $^{209}Pb_{82}$ и α-частицы, соответственно.

Из закона сохранения энергии (3.1б) получаем:

$$[M(^{213}Po_{84}) - M(^{209}Pb_{82}) - M_\alpha]c^2 = T_{Pb} + T_\alpha \qquad (3.3б)$$

Левая часть равенства (3.3б) равна энергии реакции $\boldsymbol{Q}$, поэтому

$$\boldsymbol{Q} = T_{Pb} + T_\alpha \qquad (3.4б)$$

Поскольку энергия реакции $\boldsymbol{Q}$ = 8.384 МэВ значительно меньше энергии покоя дочернего ядра и α-частицы

$$M(^{209}Pb_{82})c^2 = 194664.548 \text{ МэВ}$$

$$M_\alpha c^2 = M(^4He_2)c^2 = 3728.771 \text{ МэВ}, \qquad (3.5б)$$

кинетические энергии дочернего ядра $^{209}Pb_{82}$ и α-частицы можно определить в нерелятивистском пределе

$$T_{Pb} = \frac{\vec{P}_{Pb}^2}{2M\left(^{209}Pb_{82}\right)}, T_\alpha = \frac{\vec{P}_\alpha^2}{2M_\alpha}. \qquad (3.6б)$$

Используя далее закон сохранения импульса (3.2б), мы можем найти связь между кинетическими энергиями $T_{Pb}$ и $T_\alpha$:

$$T_\alpha = \frac{\vec{P}_{Pb}^2}{2M_\alpha} = \frac{M\left(^{209}Pb_{82}\right)}{M_\alpha}\frac{\vec{P}_{Pb}^2}{2M\left(^{209}Pb_{82}\right)} = \frac{M\left(^{209}Pb_{82}\right)}{M_\alpha}T_{Pb}, \qquad (3.7б)$$

т.е.

$$T_\alpha = \frac{M\left(^{209}Pb_{82}\right)}{M_\alpha}T_{Pb}. \qquad (3.8б)$$

Подставим соотношение (3.8б) в (3.4б) и выразим кинетическую энергию $T_{Pb}$ дочернего ядра $^{209}Pb_{82}$ через энергию, освобождаемую в реакции:

$$T_{Pb} = \frac{M_\alpha}{M\left({}^{209}Pb_{82}\right) + M_\alpha} Q . \qquad (3.9б)$$

Таким образом, доля кинетической энергии дочернего ядра $T_{Pb}$ в энергии реакции $\boldsymbol{Q}$ равна:

$$\frac{T_{Pb}}{Q} = \frac{M_\alpha}{M\left({}^{209}Pb_{82}\right) + M_\alpha} = 0.019 = 1.9\% , \qquad (3.10б)$$

где мы воспользовались численными значениями масс ядер (3.2а).

**Ответ**: Доля кинетической энергии дочернего ядра $^{209}Pb_{82}$ в энергии, выделяемой в реакции $^{213}Po_{84} \rightarrow {}^{209}Pb_{82} + \alpha$, равна 1.9%.

Решение 11.25 в. Скорость отдачи дочернего ядра $\boldsymbol{v_{Pb}}$ определим, используя связь между кинетической энергией дочернего ядра $T_{Pb}$ и скоростью $\boldsymbol{v_{Pb}}$

$$T_{Pb} = \frac{M\left({}^{209}Pb_{82}\right)}{2} v_{Pb}^2 \qquad (3.1в)$$

и кинетической энергией $T_{Pb}$ и энергией реакции $\boldsymbol{Q}$, определяемой уравнением (3.9б). Приравнивая (3.1в) к (3.9б) получим:

$$\frac{M\left({}^{209}Pb_{82}\right)}{2} v_{Pb}^2 = \frac{M_\alpha}{M\left({}^{209}Pb_{82}\right) + M_\alpha} Q . \qquad (3.2в)$$

Решаем это уравнение относительно $\boldsymbol{v_{Pb}}$:

$$v_{Pb} = \sqrt{\frac{2M_\alpha}{M\left({}^{209}Pb_{82}\right) + M_\alpha} \cdot \frac{Q}{M\left({}^{209}Pb_{82}\right)}} . \qquad (3.3в)$$

В системе СИ скорость измеряется в м/сек, т.е. $[\boldsymbol{v_{Pb}}]$ = м/сек. Для получения правильной размерности скорости $v_{Pb}$ мы должны переписать энергию реакции $\boldsymbol{Q}$ в джоулях, $[\boldsymbol{Q}]$ = Дж, а массы дочернего ядра и α-частицы в кг, $[M_x]$ = кг. Учитывая, что:

$$1 \text{ МэВ} = 1.602 \cdot 10^{-13} \text{ Дж},$$

$$1 \text{ а.е.м.} = 1.661 \cdot 10^{-27} \text{ кг}, \qquad (3.4в)$$

получаем

$$\boldsymbol{Q} = 8.384 \cdot 1.602 \cdot 10^{-13} = 1.343 \cdot 10^{-12} \text{ Дж},$$
$$M(^{209}Pb_{82}) = 202.981 \cdot 1.661 \cdot 10^{-27} = 3.471 \cdot 10^{-25} \text{ кг}$$
$$M_\alpha = M(^4He_2) = 4.003 \cdot 1.661 \cdot 10^{-27} = 6.649 \cdot 10^{-27} \text{ кг} \qquad (3.5в)$$

Подставляя численные значения энергии реакции и масс ядер (3.5в) продуктов распада $^{213}Po_{84} \rightarrow {}^{209}Pb_{82} + \alpha$ в (3.3в) находим скорость дочернего ядра $^{209}Pb_{82}$:

$$\boldsymbol{v_{Pb}} = 3.814 \cdot 10^5 \text{ м/сек.} \qquad (3.6в)$$

**Ответ**: Скорость дочернего ядра $^{209}Pb_{82}$ в реакции $^{213}Po_{84} \rightarrow {}^{209}Pb_{82} + \alpha$ равна $\boldsymbol{v_{Pb}} = 3.814 \cdot 10^5$ м/сек.

## 4. β-распады ядер

Задача 11.36. Как определяются энергии, освобождаемые при $\beta^-$-распаде и $\beta^+$-распаде и К-захвате, если известны массы материнского и дочернего атомов и масса электрона?

Указание. Ядерные реакции $\beta^-$-распада, $\beta^+$-распада и К-захвата представляют собой реакции

а) испускания электрона ($e^-$) ядром $^AX_Z$

б) испускания позитрона ($e^+$) ядром $^AX_Z$

в) захвата электрона ($e^-$) ядром $^AX_Z$:

$^AX_Z \rightarrow {}^AX_{Z+1} + e^-$, $\beta^-$-распад

$^AX_Z \rightarrow {}^AX_{Z-1} + e^+$, $\beta^+$-распад

$e^- + {}^AX_Z \rightarrow {}^AX_{Z-1}$, К-захват.

Энергия, освобождаемая в реакциях $\beta^-$-распада, $\beta^+$-распада и К-захвата, равна разности энергий покоя всех частиц начального состояния и всех частиц конечного состояния. Обратим внимание на вопрос задачи, в котором спрашивается выразить энергии, выделяемые в реакциях, через массы (или энергии покоя) материнских и дочерних атомов и электронов. Напомним, что масса позитрона равна массе электрона. Масса атома $M(^AX_Z)_{ат}$ с ядром $^AX_Z$ равна:

$$M(^A X_Z)_{ат} = M(^A X_Z) + Z\, m_e,$$

где $Z\, m_e$ - масса всех электронов. Энергией связи электронов в атоме мы пренебрегаем.

Решение 11.36 а. Энергия, выделяемая в $\beta^-$-распаде $^A X_Z \rightarrow {}^A X_{Z+1} + e^-$ равна:

$$\boldsymbol{Q_{\beta-}} = [M(^A X_Z) - M(^A X_{Z+1}) - m_e]c^2\ . \qquad (4.1а)$$

Заменим массы ядер массами атомов:

$$M(^A X_Z) = M(^A X_Z)_{ат} - Z\, m_e$$

$$M(^A X_{Z+1}) = M(^A X_{Z+1})_{ат} - (Z+1)\, m_e. \qquad (4.2а)$$

Подставим соотношения (4.2а) в (4.1а) и получим энергию, выделяемую в $\beta^-$-распаде

$$\boldsymbol{Q_{\beta-}} = [M(^A X_Z)_{ат} - Z\, m_e - M(^A X_{Z+1})_{ат} + (Z+1)\, m_e - m_e]c^2 =$$

$$= [M(^A X_Z)_{ат} - M(^A X_{Z+1})_{ат}]\, c^2. \qquad (4.3а)$$

Правую часть можно переписать следующим образом:

$$\boldsymbol{Q_{\beta-}} = (M_м - M_д)c^2, \qquad (4.4а)$$

где $M_м = M(^A X_Z)_{ат}$ и $M_д = M(^A X_{Z+1})_{ат}$ - массы материнского и дочернего атомов.

**Ответ**: Энергия, выделяемая при $\beta^-$-распаде, равна разности энергий покоя материнского и дочернего атомов: $\boldsymbol{Q_{\beta-}} = (M_м - M_д)c^2$.

Решение 11.36 б. Энергия, выделяемая в $\beta^+$-распаде $^A X_Z \rightarrow {}^A X_{Z-1} + e^+$, равна

$$\boldsymbol{Q_{\beta+}} = [M(^A X_Z) - M(^A X_{Z-1}) - m_e]c^2\ . \qquad (4.1б)$$

Заменим массы ядер массами атомов:

$$M(^A X_Z) = M(^A X_Z)_{ат} - Z\, m_e$$

$$M(^A X_{Z-1}) = M(^A X_{Z-1})_{ат} - (Z-1)\, m_e \qquad (4.2б)$$

Подставляя соотношения (4.2б) в (4.1б), получим:

$$\boldsymbol{Q_{\beta+}} = [M(^A X_Z)_{ат} - Zm_e - M(^A X_{Z-1})_{ат} + (Z-1)m_e - m_e]c^2 =$$

$$= [M(^A X_Z)_{ат} - M(^A X_{Z-1})_{ат} - 2m_e]\, c^2. \qquad (4.3б)$$

Перепишем правую часть выражения (4.3б) в виде:

$$\boldsymbol{Q}_{\beta+} = (M_м - M_д - 2m_e)c^2. \qquad (4.4б)$$

**Ответ:** Энергия, выделяемая в $\beta^+$-распаде, определяется разностью масс материнского атома и дочернего атома, и удвоенной массой электрона.

Решение 11.36 в. Энергия, выделяемая в реакции К-захвата $e^- + {}^AX_Z \rightarrow {}^AX_{Z-1}$, равна

$$\boldsymbol{Q}_K = [m_e + M({}^AX_Z) - M({}^AX_{Z-1})]c^2. \qquad (4.1в)$$

Заменим массы ядер массами атомов (см. уравнение (4.2б))

$$\boldsymbol{Q}_K = [m_e + M({}^AX_Z)_{ат} - Zm_e - M({}^AX_{Z-1})_{ат} + (Z-1)m_e]c^2 =$$

$$= [M({}^AX_Z)_{ат} - M({}^AX_{Z-1})_{ат}]\, c^2. \qquad (4.2в)$$

Таким образом, энергия, выделяемая в К-захвате, равна разности энергий покоя материнского и дочернего атомов.

$$\boldsymbol{Q}_K = (M_м - M_д)c^2. \qquad (4.3в)$$

**Ответ**: Энергия, выделяемая в К-захвате, определяется разностью энергий покоя материнского и дочернего атомов.

## 5. Ядерные реакции

Задача 13.21. Вычислить пороговую кинетическую энергию налетающей частицы в реакции $p + {}^3H_1 \rightarrow {}^3He_2 + n$, если налетающей частицей является:

а) протон

б) ядро трития

Указание. Для решения задач о вычислении пороговых кинетических энергий в ядерных реакциях

$$A + B = C + D \qquad (5.1)$$

необходимо:

1) знать, что энергия $\boldsymbol{Q}$, выделяемая в реакции, равна

$$\boldsymbol{Q} = (M_A + M_B - M_C - M_D)\, c^2, \qquad (5.2)$$

где $M_A$, $M_B$, $M_C$ и $M_D$ - массы взаимодействующих частиц.

2) использовать закон сохранения энергии и импульса:

$$M_A c^2 + T_A + M_B c^2 + T_B = M_C c^2 + T_C + M_D c^2 + T_D,$$

$$\vec{p}_A + \vec{p}_B = \vec{p}_C + \vec{p}_D, \qquad (5.3)$$

где $T_A$, $T_B$, $T_C$ и $T_D$ - кинетические энергии частиц A, B, C и D, а $\vec{p}_A$, $\vec{p}_B$, $\vec{p}_C$ и $\vec{p}_D$ - их импульсы.

При изучении реакции (5.1) в так называемой « лабораторной системе координат » одна из частиц A или B покоится. Покоящаяся частица называется мишенью. Пусть мишенью будет частица B. В этом случае $T_B = \vec{p}_B = 0$, и закон сохранения энергии и импульса принимает вид:

$$M_A c^2 + T_A + M_B c^2 = M_C c^2 + T_C + M_D c^2 + T_D,$$

$$\vec{p}_A = \vec{p}_C + \vec{p}_D. \qquad (5.4)$$

Перепишем соотношение (5.4) через энергию реакции $\boldsymbol{Q}$.

$$T_A = T_C + T_D + \boldsymbol{Q},$$

$$\vec{p}_A = \vec{p}_C + \vec{p}_D. \qquad (5.5)$$

Теперь мы можем ввести понятие пороговой кинетической энергии.

**Определение**: пороговой кинетической энергией $(T_A)_{пор}$ называют кинетическую энергию налетающей частицы, при которой дочерние частицы рождаются с нулевой кинетической энергией относительного движения в системе центра масс.

Согласно определению пороговой кинетической энергии, мы должны разложить движение дочерних ядер C и D на движение центра инерции и относительное движение. В нерелятивистском приближении, когда кинетическая энергия взаимодействующих частиц значительно меньше их энергий покоя, импульсы частиц C и D пропорциональны скоростям $\vec{v}_C$ и $\vec{v}_D$, т.е.

$$\vec{p}_C = M_C \vec{v}_C, \; \vec{p}_D = M_D \vec{v}_D. \qquad (5.6)$$

Скорость центра инерции $\vec{V}$ и относительного движения $\vec{v}$ равны:

$$\vec{V} = \frac{M_C \vec{v}_C + M_D \vec{v}_D}{M_C + M_D}, \ \vec{v} = \vec{v}_C - \vec{v}_D. \tag{5.7}$$

Легко показать с помощью простых алгебраических преобразований, что

$$\begin{aligned} T_C + T_D &= \frac{M_C \vec{v}_C^{\,2}}{2} + \frac{M_D \vec{v}_D^{\,2}}{2} = \frac{M_C + M_D}{2} \vec{V}^2 + \frac{\mu \vec{v}^2}{2} = \\ &= \frac{\vec{P}^2}{2(M_C + M_D)} + \frac{\vec{p}^2}{2\mu} \end{aligned} \tag{5.8}$$

т.е.

$$T_C + T_D = \frac{\vec{P}^2}{2(M_C + M_D)} + \frac{\vec{p}^2}{2\mu}, \tag{5.9}$$

где

$$\vec{P} = (M_C + M_D)\vec{V} = M_C \vec{v}_C + M_D \vec{v}_D = \vec{p}_C + \vec{p}_D,$$

$$\vec{p} = \mu \vec{v} = \frac{M_C M_D}{M_C + M_D}(\vec{v}_C - \vec{v}_D) \tag{5.10}$$

и

$$\mu = \frac{M_C M_D}{M_C + M_D}.$$

Напомним, что μ называется приведённой массой системы частиц C и D.

Из закона сохранения импульса, определяемого уравнением (5.5), получаем

$$\vec{P} = \vec{p}_C + \vec{p}_D = \vec{p}_A. \tag{5.11}$$

Подстановка уравнения (5.11) в уравнение (5.9) дает

$$T_C + T_D = \frac{M_A}{M_C + M_D} T_A + \frac{\vec{p}^2}{2\mu}, \tag{5.12}$$

где мы учли, что $T_A = \vec{p}_A^2 / 2M_A$ - кинетическая энергия налетающей частицы A.

С помощью уравнения (5.12) мы преобразуем уравнение (5.5) к следующему виду:

$$T_A = \frac{M_A}{M_C + M_D} T_A + \frac{\vec{p}^2}{2\mu} - Q. \tag{5.13}$$

Решение этого уравнения относительно $T_A$ дает:

$$T_A = \frac{M_C + M_D}{M_B - Q/c^2}\left(\frac{\vec{p}^2}{2\mu} - Q\right). \qquad (5.14)$$

Поскольку энергия покоя частицы-мишени значительно больше энергии, выделяемой в реакции $M_B c^2 >> |\boldsymbol{Q}|$, то выражение (5.14) может быть упрощено:

$$T_A = \frac{M_C + M_D}{M_B}\left(\frac{\vec{p}^2}{2\mu} - Q\right). \qquad (5.15)$$

Приравнивая нулю кинетическую энергию относительного движения частиц C и D, $\frac{\vec{p}^2}{2\mu} = 0$, получаем пороговую кинетическую энергию реакции (5.1):

$$T_A = \frac{M_C + M_D}{M_B}(-Q). \qquad (5.16)$$

Поскольку сумма энергий покоя начальных частиц A и B значительно больше энергии выделяемой в реакции (5.1), то можно $M_C + M_D$, заменить на $M_A + M_B$, используя равенство $M_C + M_D = M_A + M_B - \boldsymbol{Q}$ и пренебрегая $\boldsymbol{Q}$.

$$T_A = \frac{M_A + M_B}{M_B}(-Q). \qquad (5.17)$$

Перейдём к изучению реакции

$$p + {}^3H_1 \rightarrow {}^3He_2 + n \qquad (5.18)$$

Массы взаимодействующих частиц равны:

$$M_p = 1 + \Delta({}^1H_1) = 1.007825 \text{ а.е.м.}$$
$$M({}^3H_1) = 3 + \Delta({}^3H_1) = 3.016005 \text{ а.е.м.}$$
$$M_n = 1 + \Delta({}^1n_0) = 1.008665 \text{ а.е.м.} \qquad (5.19)$$
$$M({}^3He_2) = 3 + \Delta({}^3He_2) = 3.016030 \text{ а.е.м.}$$

Энергия реакции (5.18), равна:

$$\boldsymbol{Q} = [M_p + M({}^3H_1) - M({}^3He_2) - M_n]c^2 =$$
$$= -0.000865 \cdot 931.494 = -0.806 \text{ МэВ.}$$

Решение 13.21 а. Если частицей мишенью является тритий (${}^3H_1$), то пороговая кинетическая энергия реакции (5.18) равна:

$$\left(T_p\right)_{пор} = \frac{M_p + M\left({}^3H_1\right)}{M\left({}^3H_1\right)}\left(-Q\right) = 1.075 МэВ \,.$$

**Ответ**: Пороговая кинетическая энергия протона в реакции p + ${}^3H_1 \rightarrow {}^3He_2$ + n равна $(T_p)_{пор}$ = 1.075 МэВ.

<u>Решение 13.21 б.</u> Если мишенью является протон, то пороговая кинетическая энергия реакции (5.28) равна

$$\left(T_{{}^3H_1}\right)_{пор} = \frac{M_p + M\left({}^3H_1\right)}{M_p}\left(-Q\right) = 3.218 МэВ$$

**Ответ**: Пороговая кинетическая энергия в реакции p + ${}^3H_1 \rightarrow {}^3He_2$ + n равна $(T({}^3H_1))_{пор}$ = 3.218 МэВ.

## 6. Энергетические уровни ядер

Задача 13.45. Мишень $^{7}Li_3$ бомбардируют пучком нейтронов с кинетической энергией $T_0$ = 1,00 МэВ. Определить энергию возбуждения ядер, возникающих в результате неупругого рассеяния нейтронов, если энергия нейтронов, рассеянных под прямым углом к падающему пучку, равна T = 0.33 МэВ.

Указание. Определим сначала понятия **упругого** и **неупругого** рассеяния.

**Упругим** называют рассеяние частиц A и B, если состав частиц начального состояния совпадает с составом частиц конечного состояния

$A + B = A + B$.

**Неупругим** называют рассеяние частиц A и B, если состав частиц начального состояния отличается от состава частиц конечного состояния (внутреннее строение, превращение в другие частицы или дополнительное рождение новых частиц)

$A + B = C + D$.

Рассмотрим неупругое рассеяние, когда частицы A и D совпадают, а частица (ядро) C является возбуждённым состоянием частицы (ядра) B, которую обозначают как $C = B^*$.

$A + B = B^* + A$.

Если $M_B$ – масса ядра B, то масса возбуждённого ядра $M_{B^*}$ равна:

$M_{B^*} = M_B + E_B^*/c^2$,

где $E_B^*$ - энергия возбуждения.

Вычисление энергии возбуждения $E_B^*$ в реакции $A + B = B^* + A$ осуществляется с использованием закона сохранения энергии и импульса.

Решение 13.45 По условию задачи, в результате неупругого рассеяния нейтронов на ядрах $^{7}Li_3$ образуются ядра возбуждённого состояния - $^{7}Li_3^*$.

Ядерная реакция имеет вид:

$$n + {}^7Li_3 \rightarrow {}^7Li_3^* + n. \quad (6.1)$$

Если $M({}^7Li_3)$ – масса ${}^7Li_3$, то масса возбуждённого ядра ${}^7Li_3^*$ равна:

$$M({}^7Li_3^*) = M({}^7Li_3) + E^*/c^2 , \quad (6.2)$$

где $E^*$ - энергия возбуждения.

По условию задачи ядро ${}^7Li_3$ является мишенью и покоится. Закон сохранения энергии и импульса имеет в этом случае следующий вид:

$$M_nc^2 + T_0 + M({}^7Li_3)c^2 = M({}^7Li_3^*)c^2 + T({}^7Li_3^*) + M_nc^2 + T,$$

$$\vec{p}_0 = \vec{p} + \vec{p}\left({}^7Li_3^*\right), \quad (6.3)$$

где $T({}^7Li_3^*)$ и $\vec{p}\left({}^7Li_3^*\right)$ - кинетическая энергия и импульс возбуждённого лития, $M_n$ – масса нейтрона, а $\vec{p}$ - импульс конечного нейтрона.

Закон сохранения энергии перепишем с учётом (6.2):

$$M_nc^2 + T_0 + M({}^7Li_3)c^2 = M({}^7Li_3)c^2 + E^* + T({}^7Li_3^*) + M_nc^2 + T. \quad (6.4)$$

Решая это уравнение относительно $E^*$, получим:

$$E^* = T_0 - T - T({}^7Li_3^*). \quad (6.5)$$

Соотношение (6.5) представляет собой линейное уравнение с одним неизвестным – кинетической энергией возбуждённого лития $T({}^7Li_3^*)$, которую можно определить, используя закон сохранения импульса:

$$\vec{p}_0 = \vec{p} + \vec{p}\left({}^7Li_3^*\right) \quad (6.6В$$

Кинетическая энергия возбуждённого лития равна:

$$T\left({}^7Li_3^*\right) = \frac{\vec{p}^2\left({}^7Li_3^*\right)}{2M\left({}^7Li_3^*\right)}. \quad (6.7)$$

Из уравнения (6.6) выражаем импульс возбуждённого лития $\vec{p}\left({}^7Li_3^*\right)$ через импульсы нейтронов:

$$\vec{p}\left({}^7Li_3^*\right) = \vec{p}_0 - \vec{p}. \quad (6.8)$$

Подставляя уравнение (6.8) в уравнение (6.7) находим кинетическую энергию возбуждённого лития $T({}^7Li_3^*)$:

$$T\left({}^7Li_3^*\right)=\frac{\left(\vec{p}_0-\vec{p}\right)^2}{2M\left({}^7Li_3^*\right)}=\frac{1}{2M\left({}^7Li_3^*\right)}\left(\vec{p}_0^{\,2}+\vec{p}^{\,2}-2\vec{p}_0\vec{p}\right). \qquad (6.9)$$

Поскольку по условию задачи импульсы начального и конечного нейтронов ортогональны, $\vec{p}_0\vec{p}=0$, то уравнение (6.9) приводится к виду:

$$\begin{aligned}T\left({}^7Li_3^*\right)&=\frac{\vec{p}_0^{\,2}}{2M\left({}^7Li_3^*\right)}+\frac{\vec{p}^{\,2}}{2M\left({}^7Li_3^*\right)}=\\&=\frac{M_n}{2M\left({}^7Li_3^*\right)}T_0+\frac{M_n}{2M\left({}^7Li_3^*\right)}T\end{aligned} \qquad (6.10)$$

где мы учли, что $T_0={p_0^2}/{2M_n}$ и $T={p^2}/{2M_n}$.

Подставляя уравнение (6.10) в уравнение (6.5) мы приходим к следующему соотношению

$$E^*=\left(1-\frac{M_n}{M\left({}^7Li_3\right)+E^*/c^2}\right)T_0-\left(1+\frac{M_n}{M\left({}^7Li_3\right)+E^*/c^2}\right)T, \qquad (6.11)$$

которое представляет собой алгебраическое уравнение второго порядка относительно энергии возбуждения лития ${}^7Li_3^*$.

Решение этого уравнения может быть упрощено, если принять во внимание, что

$$M({}^7Li_3)c^2 >> |E^*|. \qquad (6.12)$$

Кроме того, уравнение (6.11) может быть также упрощено, если использовать приближённые значения для массы нейтрона и массы лития

$$M_n = 1 + \Delta({}^1n_0) \approx 1 \text{ а.е.м.}$$

$$M({}^7Li_3) = 7 + \Delta\,({}^7Li_3) \approx 7 \text{ а.е.м.} \qquad (6.13)$$

Напомним, что

$$\Delta({}^1n_0) = 0.008665 \text{ а.е.м.}$$

$$\Delta\,({}^7Li_3) = 0.016005 \text{ а.е.м.} \qquad (6.14)$$

(см. таблицу 7, стр. 205 [1]).

В результате всех упрощений алгебраическое уравнение (6.11) линеаризуется относительно энергии возбуждения лития $E^*$.

Решение линеаризованного уравнения дает следующее выражение для энергии возбуждения лития $E^*$ в реакции $n + {}^7Li_3 \rightarrow {}^7Li_3^* + n$:

$$E^* = \frac{6}{7}T_0 - \frac{8}{7}T = 0.48 МэВ$$

**Ответ**: Энергия возбуждения лития в реакции $n + {}^7Li_3 \rightarrow {}^7Li_3^* + n$ равна $E^* = 0.48 МэВ$ .

## 7. Физические основы ядерной энергетики

Задача 15.2. Найти полный поток антинейтрино и уносимую им мощность из реактора с тепловой мощность 20 МВт, считая, что на каждое деление приходится пять $\beta^-$-распадов осколков, для которых суммарная энергия антинейтрино составляет около 11 МэВ.

В ядерных реакторах используют уран-235, ${}^{235}U_{92}$. Известно, что энергия, освобождающаяся в одном акте деления ядра урана равна $\boldsymbol{Q_{\beta-}}$ = 200 МэВ (см. стр. 106 [1]).

Решение 15.2. Тепловая мощность реактора P = 20 МВт = $20 \cdot 10^6$ Дж/сек определяет полную энергию, освобождающуюся при делении урана-235 в реакторе в одну секунду. Количество актов деления в одну секунду равно отношению P/$\boldsymbol{Q_{\beta-}}$. В свою очередь, количество $\beta^-$-распадов осколков в одну секунду равно:

$$\Phi = 5\frac{P}{Q_{\beta^-}} = \frac{5 \cdot 20 \cdot 10^6 \, Дж/сек}{200 \cdot 1.602 \cdot 10^{-13} \, Дж} = 3.12 \cdot 10^{18} 1/сек \, . \qquad (7.1)$$

Величина Ф определяет также полный поток антинейтрино, образованных в результате $\beta^-$-распадов осколков урана-235.

Энергия, уносимая одним антинейтрино равна E/5, где E = 11 МэВ, поэтому для мощности потока антинейтрино получаем:

$$P_{\tilde{\nu}_e} = \frac{E}{5}\Phi = \frac{1}{5} \cdot 11 \cdot 1.602 \cdot 10^{-13} \cdot 3 \cdot 12 \cdot 10^{18} = 1.1 МВт \qquad (7.2)$$

**Ответ**: Полный поток антинейтрино, образованный от $\beta^-$-распадов осколков урана в реакторе, равен Ф = 3.12·$10^{18}$ 1/сек. При суммарной энергии антинейтрино, образованных в одном акте деления Е = 11 МэВ, мощность потока антинейтрино равна $P_{\bar{\nu}_e} = 1.1 MBm$.

**Литература**